\documentclass[manuscript]{aastex}

\shorttitle{{\it Kepler} Astrometry}
\shortauthors{Monet et al.}

\begin{document}


\title{Preliminary Astrometric Results from {\it Kepler}}


\author{David G. Monet}
\affil{U. S. Naval Observatory, Flagstaff, AZ 86001}
\author{Jon M. Jenkins}
\affil{SETI Institute, Mountain View, CA 94043}
\author{Edward Dunham}
\affil{Lowell Observatory, Flagstaff, AZ 86001}
\author{Stephen T. Bryson}
\affil{NASA/Ames Research Center, Moffett Field, CA 94035}
\author{Ronald L. Gilliland}
\affil{Space Telescope Science Institute, Baltimore, MD 21218}
\author{David W. Latham}
\affil{Harvard-Smithsonian Center for Astrophysics, Cambridge MA 02138}
\author{William J. Borucki}
\affil{NASA/Ames Research Center, Moffett Field, CA 94035}
\and
\author{David G. Koch}
\affil{NASA/Ames Research Center, Moffett Field, CA 94035}


\begin{abstract}
Although not designed as an astrometric instrument, {\it Kepler} is
expected to produce astrometric results of a quality appropriate to support
many of the astrophysical investigations enabled by its photometric results.
On the basis of data collected during the first few months of operation,
the astrometric precision for a single 30 minute measure
appears to be better than 4 milliarcseconds (0.001 pixel).
Solutions for stellar parallax and proper
motions await more observations, but the analysis of the
astrometric residuals from a local solution in the vicinity of a
star have already proved to be
an important tool in the process of confirming the
hypothesis of a planetary transit.
\end{abstract}





\section{Introduction}

The measurement of astrometric parameters, particularly the parallax, for
{\it Kepler} stars is a critical component of computing the physical
values for various stellar parameters using the relative values that are
computed from the photometric analysis.  If the {\it Kepler} data can
be shown to have the necessary astrometric accuracy, then such a conversion
can be included in the processing for most if not all {\it Kepler} stars.
Although the discussion of {\it Kepler} astrometric accuracy must await
more data and modeling, the very high precision of {\it Kepler} positions
is already a powerful tool for understanding the photometric variations
of stars and the possible presence of planetary companions.
Detailed discussions of the {\it Kepler} spacecraft and mission are presented
by \citet{bor10} and \citet{koc10}.  A short overview for
the astrometric discussion is the following.  The Schmidt telescope has a 1.4-m
primary and a 0.95-m corrector,
and the photometer is a mosaic of 42 charge coupled devices (CCDs).  The
boresight of the telescope remains constant for the mission, but the spacecraft
rolls 90 degrees every three months.  Due to restrictions in memory and
bandwidth, only the pixels associated with the target stars are sent
to the ground for processing.  Target
stars are defined by software, and pixels not associated with targets are saved
only infrequently.  The basic integration time is 6.02 seconds, and the
long cadence (LC) sequence \citep{jen10} co-adds the target pixels for 29.4
minutes.  The co-added pixel data are sent to the ground every month for
processing and analysis.  This strategy enables the extremely high
signal-to-noise (SNR)
observations needed to achieve the planetary detection mission.

To obtain the large field of view, the image sampling is very coarse
compared to other astrometric assets.  The Pixel Response Function (PRF) is
described in great detail by \citet{bry10}, but a quick summary is as
follows.  The images contain three components, a sharp spike in the middle
that comes from optical diffraction (about 0.1 arcsecond), a wider
component with a characteristic size of 5 or 6 arcseconds set
primarily by mechanical alignment tolerances of the CCD mosaic, and
a much broader scattering profile.  The intermediate component of the
image profile produces the majority of the astrometric signal, and it
contains about 70\% of the light.
The 43 days of data available so far
\footnote{As described more fully by \citet{cal10}, the spacecraft data
are grouped by observing quarters as defined by the mandatory rolls of
the spacecraft itself.  So far, data from Quarters 0 and 1 are
available on the ground.}
have shown a remarkable astrometric precision, but the demonstration of
astrometric accuracy is still a work in progress.  Even if the centroiding
process was fully
understood, the short interval of available data precludes the lifting
of the degeneracies between effects of proper motion, parallax, and velocity
aberration.  No measured astrometric parameters are given here.  Indeed, the
entire range of observed image motion is about 0.2 pixels, and this is
dominated by the spacecraft guiding precision.

Various authors, including \citet{kin83} and \citet{kai00}, have developed
theoretical expectations for the astrometric precision of an image.
A simple approximation is
\begin{equation}
precision = FWHM/(2*SNR)
\end{equation}
where FWHM is the image full width at half maximum, and SNR is the
photometric signal-to-noise
ratio of that star image.  The differences in the theoretical derivations
concern the exact value for which the approximate value of 2 is used above.
The observational confirmation of this relationship has yet to be done,
but essentially all ground- and space-based astrometric studies have
demonstrated the validity of the scaling of this relationship.  Improved
astrometric precision is obtained for smaller image FWHM, higher SNR, or
both assuming that adequate image sampling is available.

{\it Kepler} operates in a heretofore unstudied astrometric domain.  The
pixels are very large, 3.98 arcseconds, as compared to other ground-
and space-based astrometric assets, and the observed FWHM is approximately
5 to 6 arcseconds and depends on the location in the field of view.
(See \citet{bry10} for further discussion and examples.)
The effects of undersampled image components are not
captured by Eq. 1.  However, {\it Kepler} was designed for extremely high
SNR observations.  The well capacity of the 27-micron CCD pixels is
more than a million electrons, and most of the stars are bright.
As more fully discussed by \citet{cal10}, the onset of saturation in the
basic 6.02 second integration cycle is near the magnitude
{\it Kp} = 11.3. \footnote{The {\it Kepler} magnitude {\it Kp} includes a very wide
passband and is similar to an astronomical {\it R} magnitude
in central wavelength.}
A single
LC co-addition produces a SNR of about 10,000 for an bright, unsaturated,
uncrowded star.
There is no atmosphere to degrade the image quality, and
the flux is so large that effects such as sensor readout noise and dark
current are unimportant for the brightest 2-3 magnitudes of unsaturated stars.

\section{Preliminary Astrometric Investigations}

The astrometric processing of {\it Kepler} data is conceptually no different
than the traditional differential astrometric process of data from other
ground- and space-based assets.  There is no need to worry about the
actual coordinates (i.e., J2000 RA and Dec) of the stars.  Essentially
all {\it Kepler} stars are in the 2MASS catalog \citep{skr06}, and most are
in the UCAC-2 catalog \citep{zac04}.
Rather, the goal of the analysis is to measure the small
changes in position associated with proper motion, parallax, perturbations
from unseen companions, and blending with photometrically variable stars.
The current astrometric
pipeline involves three distinct steps: centroids are computed from the
pixel data, transformations are computed from each channel
\footnote{As described more fully by \citet{jen10}, each CCD is split into
two channels by the flight electronics.  The astrometric verification of
the stability between the two channels of a single CCD has yet to
be performed.}
of each LC
co-addition into an intermediate coordinate system, and solutions for each star
are computed using the intermediate coordinates and terms such as
time, parallax factor, etc.
Steps two and three are iterated a few times, and convergence
is quite rapid.  Details of each of these steps are presented in the
following subsections.

\subsection{Centroids}

Most modern centroiding algorithms fall into three classes: moment
analysis, fits to analytic functions, and fits to the instrumental
point spread function (PSF).  So far, various algorithms from the first
two classes have been implemented and tested.  The data processing
pipeline of the Science Operations Center (SOC; \citet{jen10}) compute
flux-weighted means for all stars and Gaussian fits and PSF fits for
a few stars.  PSF fitting of all stars remains a task for the future.
In a separate effort, several other centroiding algorithms based on
fitting the images to analytic functions have been evaluated, and centroids
for all stars have been computed for many of these.
The choice of centroiding algorithm requires special attention because
of the properties of the {\it Kepler} images discussed above.  The
effect of the undersampled components of the images has not been
fully evaluated, and the
number of pixels for each star has been minimized by the Optimal
Aperture algorithm \citep{bry10} so that the maximum number of stars
can be transmitted in the fixed spacecraft bandwidth.
Because the best astrometric centroiding algorithm has not been identified yet,
analysis is proceeding with parallel tracks to evaluate a few of the most
promising algorithms.

\newcommand{\bsra}{\ensuremath{19^{\mathrm{h}}22^{\mathrm{m}}40^{\mathrm{s}}}}
\newcommand{\bsdec}{{\ensuremath{+44^{\circ}30'}}}
\subsection{Transformation Coefficients}

The current astrometric pipeline supports two different coordinate systems.
For some investigations, working in a sky-based system seems appropriate.
The tangent point is taken to be the nominal {\it Kepler} boresight
($\alpha=\bsra, \delta=\bsdec$, J2000)
and a simple tangent plane projection
is computed from the nominal positions of the stars listed in the {\it Kepler}
Input Catalog (KIC).
\footnote{http://archive.stsci.edu/kepler/kepler\_fov/search.php}
In this coordinate system, effects such as differential velocity
aberration and parallax are easy to visualize.  For other investigations,
a channel-based coordinate system seems appropriate, and effects arising
from the structure and behavior of CCD pixels are easier to visualize.  Of
particular importance are effects based on where the star falls with
respect to the pixel grid, a term called ``pixel phase''.  In
either coordinate system, a separate set of transformation coefficients
is computed for the measures from each channel for each cadence.
A sample of 1000 known distant giant stars
was included in the {\it Kepler} star list, and are used during
the first iteration to generate a transformed coordinate system that is
as close to inertial as possible.

\begin{figure}
\epsscale{0.4}
\plotone{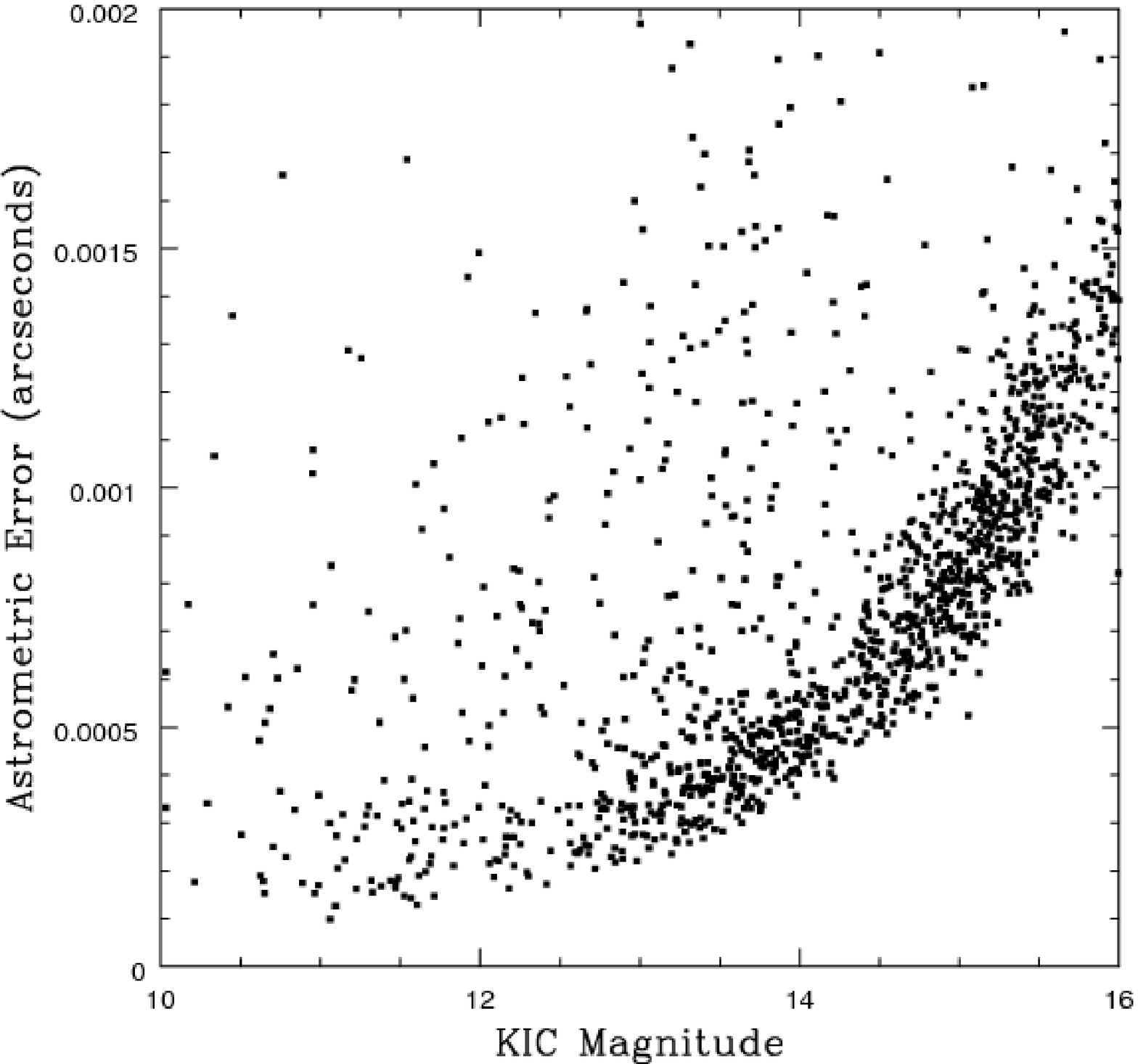}
\caption{Astrometric error as a function of {\it Kp} magnitude for
stars on Channel 2 in the Quarter 1 data collection.}
\end{figure}

\subsection{Astrometric Coefficients}

The tasks of computing the centroids and the transformation coefficients
for the {\it Kepler} data are similar to their
counterparts for traditional ground- and space-based assets.  It is the
modeling of the measured positions for each star that involves special
attention, and this flows from the extremely high SNR of the data.
The simplest solutions based on
computing only the mean positions from the data currently available
show an astrometric precision near 20 milliarcseconds (about 0.005 pixels).
This value and those presented below refer to the uncertainty in a
single measure of a single axis (row or column) for a single star from a
single LC co-addition.  Because these solutions are local and contain only a
small number of measures, they should be construed as estimators of the
astrometric precision and not of the overall astrometric accuracy of
the spacecraft and photometer.
Adding terms that model the differential velocity aberration across the
{\it Kepler} field reduces this error to about 4 milliarcseconds (about
0.001 pixels), and adding terms arising from the pixel phase reduce the
errors to about 2 milliarcseconds (about 0.0005 pixels).  These pixel
phase terms are empirical fits, and are not derived from detailed
modeling of the image formation and sampling processes.

During the first 33.5 days of science operations that followed the end of
commissioning, the number of stars was increased to 156,000.  The volume
of data for this number of stars observed every 29.4 minutes is large,  and the
data from the 84 channels are diverse.  Thus it is difficult to characterize
the entirety of the astrometric solution with a single number.
Again, much development in the astrometric processing is needed because
every {\it Kepler} star is important.  Although only preliminary
measures of astrometric precision have been obtained, it is reassuring
to see that the observed errors follow the prediction based on the
flux of the stars involved.  Fig. 1 shows the errors for stars
in a single channel as a function of the measured {\it Kp},
and demonstrates that the error rises as the SNR decreases.

\section{Local Astrometric Solutions and the ``Rain''  Plots}

A special case of the astrometric solutions described above can be computed
in the vicinity of individual stars.  Under the assumptions of small
parallax and adequate removal of differential velocity aberration, the
equations for the apparent place of a star can be linearized.  Simple trend
analysis produces a robust estimator for the mean position of a star, and
residuals from each measurement are computed.  This enables astrometric
processing to contribute to the understanding of the {\it Kepler} stars.
As more fully discussed by \citet{bah10}, what appears to be a single
object can be two or more stars, and each can have photometric variability.
Such astrations can mimic the photometric properties of a transiting
planet, and adding astrometry to the vetting procedure can assist in the
confirmation or denial process.  Fig. 2 shows astrometric residuals
for two stars.
The upper star is a blend of a variable star and one or more
constant stars while the lower shows residuals that are typical for a bright,
constant star.
The astrometric amplitude of the variable star is huge - almost 0.02 pixels.

The astrometric behavior of an image composed of an unknown number
of stars each of which having unknown photometric variations is complicated.
However, the simple case of two stars, a small photometric variation,
and centroids computed from flux-weighted means provides much insight.
Where $\Delta s$ is the true separation of the stars, $\delta s$ is the
small measured
astrometric shift, $F$ is the small relative brightness of the fainter
B component compared to the brighter A component, and $f$ is the small
relative
change in total brightness due to a transit or stellar variability, the
observed astrometric shift assuming that the A component is variable is
given by
\begin{equation}
\delta s = F f \Delta s
\end{equation}
If the B component has the photometric variation, the observed shift is
\begin{equation}
\delta s = f \Delta s
\end{equation}
When the {\it Kepler} observations of $\delta s$ and $f$ are combined
with high resolution imagery that can measure $\Delta s$, $F$, and
other characteristics such as stellar colors, then the model of the
composite image can be improved.

On the basis of its photometric signature alone, the {\it Kepler}
Object of Interest (KOI-) 15 might
involve a transiting planet.  The analysis of the combined photometric
and astrometric residuals denies this hypothesis.
Fig. 3. shows the time series
astrometric and photometric residuals for KOI-15
after they have been high-pass filtered so as to emphasize signals with
shorter timescales.  A different visualization of the same data is called
a ``Rain Plot'', and is shown in Fig. 4.  Clearly, the astrometric and
photometric residuals for KOI-15 are strongly correlated, and these
correlations are the signature of an astration that includes one or more
relatively constant stars and a background eclipsing binary.  Indeed,
the secondary eclipse is more apparent in the astrometric residuals
than in the photometric residuals.  A true transiting planet system
should not show these correlations.  As suggested by the Rain Plot,
the pixel data were re-examined and the offending variable star was
identified as being about 11 arcseconds away from and about 4.8 {\it Kp}
magnitudes fainter than the brighter star.

\section{Conclusions}

Although based on just a small fraction of the data expected from the
entire mission, the following conclusions can be drawn.

a) Both the preliminary version of the full astrometric solution and the
locally linearized astrometric solution indicate that the precision of a
single measure of a typical star is about 0.004 arcsecond (about 0.001 pixel).
Results such as those shown in Figs. 2 and 3 suggest that the precision
for some stars might be substantially better.

b) On the basis of the limited data available so far, many astrometric
effects cannot be separated.  The astrometric precision should enable
the measurement of the proper motions of the known large motion stars
in the star list, but as yet proper motion is indistinguishable from
differential velocity aberration, pixel phase, and similar effects.

c) Because of the large pixel size, {\it Kepler} images can often be
blends of multiple stars.  If one or more components of such a blend is
a photometrically variable star, the astrometric position of the image
can have a significant motion.  Tools such as the Rain Plot have
already demonstrated the utility of combining astrometric and photometric
processing in the evaluation of planetary transit candidates.

d) Whereas the astrometric precision of {\it Kepler} data has been
demonstrated, the astrometric accuracy has yet to be evaluated.

In summary, it is the extremely high SNR of the {\it Kepler} photometer
that enables the astrometric analysis of {\it Kepler} data.  Even with
just the preliminary data, astrometric analysis is providing an important
tool for the physical understanding of the observations.  Should the astrometric
results continue to follow the theoretical expectation of improving with
the SNR, then solutions for the parallax and proper motions of all stars
will be computed.

\begin{figure}
\epsscale{0.4}
\plotone{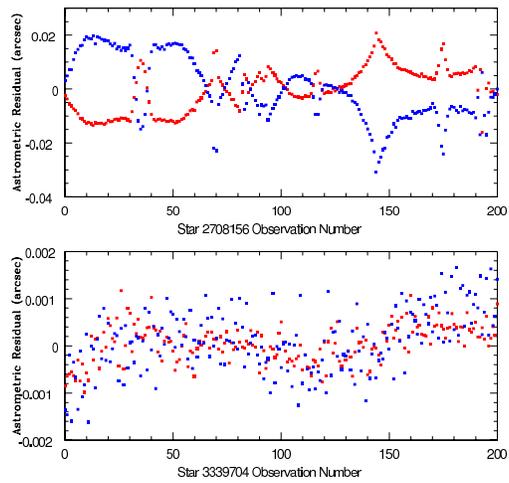}
\caption{Astrometric residuals from a blended variable (top) and a non-variable
star (bottom) taken from a solution for a single channel.
Red symbols are from the columns
and blue symbols are from the rows.}
\end{figure}

\begin{figure}
\epsscale{0.4}
\plotone{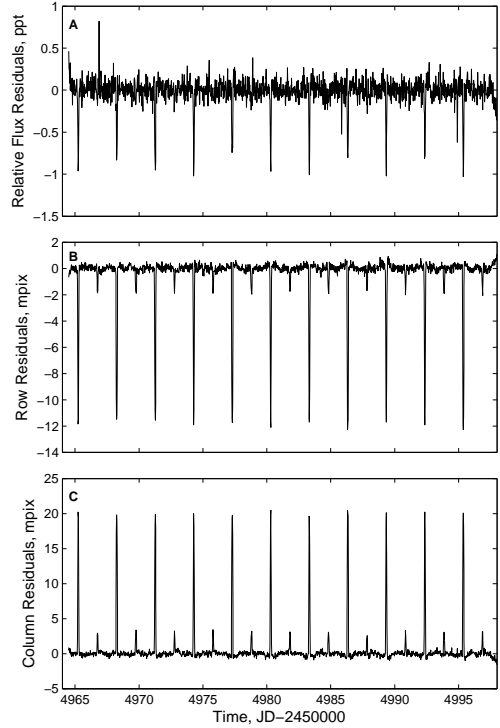}
\caption{Time series residuals for the relative photometric flux (A) in parts
per thousand, and the relative astrometric row (B) and column (C) residuals
in millipixels (1 pixel = 3.98 arcseconds) for KOI-15.  The observations
have been filtered to remove long-period trends.  The strong correlation
between these residuals is taken as evidence that this {\it Kepler} star
is actually an astration of one or more relatively constant stars and a
background eclipsing binary.}
\end{figure}

\begin{figure}
\epsscale{0.4}
\plotone{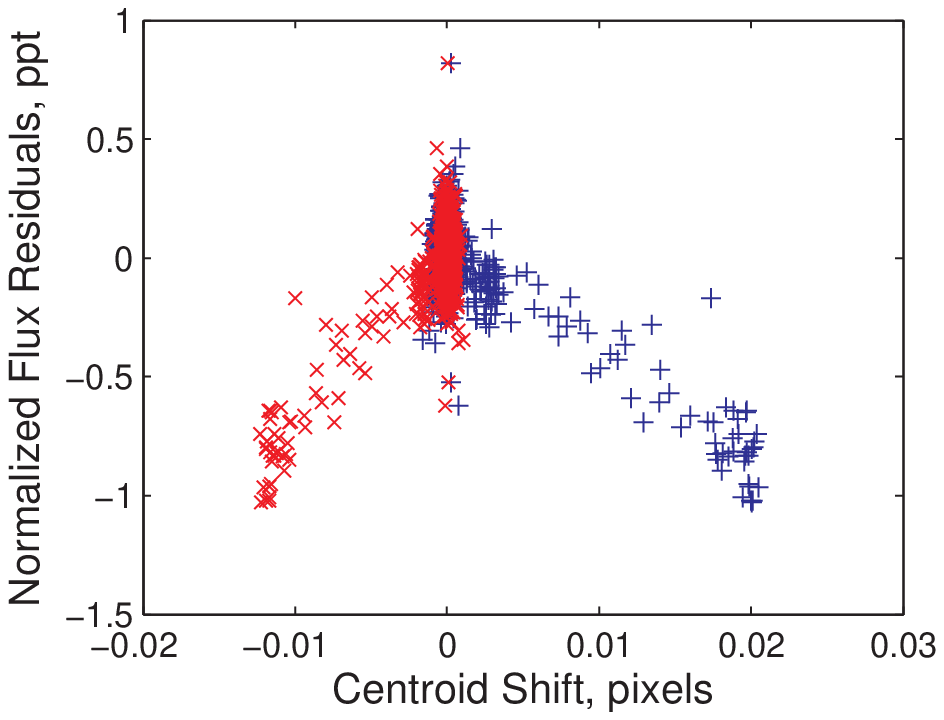}
\caption{The Rain Plot for KOI-15.  This visualization shows the correlation
between the photometric and astrometric row (red X) and column (blue +)
residuals.
The strong correlation between the position and brightness is evidence that
this is an astration of one or more relatively constant stars and
a background eclipsing binary.}
\end{figure}

\acknowledgments

Funding for this Discovery mission is provided by NASA's Science Mission
Directorate.
Many people have contributed to the success of the {\it Kepler} Mission, and
the authors wish to express their profound thanks to all.


\it{Facilities:} \facility{The Kepler Mission}.

\end{document}